\documentclass[10pt, twocolumn]{IEEEtran}
\IEEEoverridecommandlockouts
\usepackage{graphicx}
\usepackage{subfigure}
\usepackage{color}
\usepackage{epsfig}
\usepackage{amssymb}
\usepackage{amsmath}
\usepackage{amsthm}
\usepackage{latexsym}
\usepackage{setspace}

\newcommand{\bPr}[1]{{\Pr}\left(#1\right)}

\newcommand{\bP}[2]{{P}_{#1}\left({#2}\right)}

\newcommand{\cX}{{\mathcal X}}
\newcommand{\cJ}{{\mathcal J}}

\newcommand{\cY}{{\mathcal Y}}

\newcommand{\cU}{{\mathcal U}}

\newcommand{\cM}{{\mathcal M}}
\newcommand{\cA}{{\mathcal A}}
\newcommand{\cL}{{\mathcal L}}

\newcommand{\cV}{{\mathcal V}}

\newcommand{\cR}{{\mathcal R}}

\newcommand{\ep}{\epsilon}
\newcommand{\la}{\lambda}

\newtheorem{theorem}{Theorem}

\newtheorem{corollary}[theorem]{Corollary}

\newtheorem{lemma}[theorem]{Lemma}

\theoremstyle{remark}
\newtheorem*{remark*}{Remark}
\newtheorem*{remarks*}{Remarks}

\newtheorem{example}{Example}
\theoremstyle{definition}
\newtheorem{definition}{Definition}

\begin{document}
\IEEEoverridecommandlockouts

\title{Distributed Function Computation with Confidentiality}

\author{
\IEEEauthorblockN{Himanshu Tyagi$^\dag$} }

\maketitle {\renewcommand{\thefootnote}{}\footnotetext{

\vspace{.02in}\noindent This work was supported by the U.S.
National Science Foundation under Grants CCF0830697 and
CCF1117546. The material in this paper was presented in part 
at the 2012 IEEE International Symposium on Information Theory. 

\noindent$^\dag$Department of Electrical and Computer Engineering,
 and Institute for Systems Research, University of Maryland, College
 Park, MD 20742, USA. Email: tyagi@umd.edu.
}}

\renewcommand{\thefootnote}{\arabic{footnote}}
\setcounter{footnote}{0}

\begin{abstract}
A set of terminals observe correlated data and seek to compute
functions of the data using interactive public communication. At
the same time, it is required that the value of a private function
of the data remains concealed from an eavesdropper observing this communication. 
In general, the private function and the functions
computed by the nodes can be all different. We show that a class
of functions are securely computable if and only if the
conditional entropy of data given the value of private function is
greater than the least rate of interactive communication required
for a related multiterminal source-coding task. A
single-letter formula is provided for this rate in special cases.
\end{abstract}


\section{Introduction}\label{s_int}
We consider the following distributed function computation problem with a confidentiality
requirement. The terminals in a set $\cM = \{1, ..., m\}$ observe
correlated data, and wish to compute functions $g_1, ..., g_m$, respectively, of their
collective data. To this end, they communicate
interactively over a noiseless channel of unlimited capacity. It
is required that this communication must not reveal the value
of a specified private function $g_0$ of the data. If such a
communication protocol exists, the functions $g_0, g_1, ..., g_m$
are said to be {\it
securely computable}. 
We formulate a Shannon theoretic multiterminal source model that
addresses the basic question: \emph{When are the functions $g_0, g_1, ..., g_m$ securely computable?}
%

Applications of this formulation include distributed 
computing over public communication networks and function
computation over sensor networks in hostile environments. 
In contrast to the classic notion of 
secure computing in cryptography \cite{Yao82}, 
we assume that the nodes are trustworthy but their public 
communication network can be accessed by an eavesdropper. 
We examine the 
feasibility of certain distributed computing tasks without revealing 
a critical portion of the data to the eavesdropper; the function $g_i$, $i =1, ..., m$, denotes
the computation requirements of the $i$th terminal, while the critical data is represented by the 
value of private function $g_0$. As an example, consider a data download problem 
in a sensor network. The central server terminal $1$ downloads binary data
from terminals $2, ..., m$, while the latter terminals compute the symbolwise maxima. An
observer of the communication must not learn of the data of terminals $2, ..., m$.

The answer to the general question above remains open. The 
simplest case of interest
when the terminals in a subset $\cA$ of $\cM$ compute only the private
function $g_0$ and those not in $\cA$ perform no computation was
introduced in \cite{Tya11}. The data download problem, upon dropping 
the computation requirements for terminals $2, ..., m$, reduces to this setting.
It was shown 
that if $g_0$ is securely computable (by the terminals in $\cA$),
then
\begin{align}\label{e_f1}
H\left(X_\cM| G_0\right) = H\left(X_\cM \right) - H\left(G_0\right)\geq R^*,
\end{align}
and $g_0$ is securely computable if
\begin{align}\label{e_f2}
H\left(X_\cM| G_0\right) > R^*,
\end{align}
where $R^*$ has the operational significance of being the minimum overall rate of
communication needed for a specific multiterminal source-coding task that
necessitates the recovery of entire data at all the terminals in
$\cA$; this task does not involve any security constraint (see Section \ref{s_formulation}
for a detailed discussion). Loosely speaking, denoting the collective data of the
terminals by the random variable (rv) $X_\cM$ and the random
value of the function $g_0$ by the rv $G_0$, the maximum rate of randomness (in the data) that is
independent of $G_0$ is $H\left(X_\cM| G_0\right)$. The conditions above
imply, in effect, that $g_0$ is securely computable if and only if this residual 
randomness 
of rate
$H\left(X_\cM| G_0\right)$ contains an interactive communication, of rate $R^*$,
for the mentioned source-coding task.

In this paper, for a broad
class of settings involving the secure computation of multiple functions, we 
establish necessary and sufficient conditions for secure computation
of the same form as (\ref{e_f1}) and (\ref{e_f2}), respectively. 
The rate $R^*$ now corresponds to, roughly, the minimum overall rate of communication 
that allows each terminal to:
\begin{itemize}
\item[(i)] accomplish its required computation task, and,

\item[(ii)] along with the private function value, recover the entire data.
\end{itemize}
This characterization of secure computability is obtained
via a general heuristic principle that leads to new results 
and further explains the results of \cite{Tya11} in a broader 
context. 

Using the sufficient condition (\ref{e_f2}), we present a specific
secure computing protocol in Section \ref{s_4} with 
a communication of rate $R^*$.
Under (\ref{e_f2}), 
the secure computing scheme in
\cite{Tya11} recovered the entire data, i.e., the collective
observations of all the terminals, at the (function seeking) terminals
in $\cA$ using communication that is independent of $G_0$. 
In fact, we observe that this is a special case of the following more general principle:
a terminal that computes the private function 
$g_0$, may recover the entire data without affecting the conditions for
secure computability. 

Unlike \cite{Tya11}, we do not provide a single-letter formula for the quantity $R^*$, 
in general; nevertheless, conditions
(\ref{e_f1}) and (\ref{e_f2}) 
provide a structural
characterization of securely computable functions in a broader setting,
generalizing the results in \cite{Tya11}. A general recipe for 
single-letter characterization
is presented which, in Example \ref{ex_1} and Corollary \ref{c_1} below, yields 
single-letter results that are new and cannot be obtained from the analysis in \cite{Tya11}.
To the best of our knowledge, the general analysis presented here is the only
known method to prove the necessity of the single-letter conditions for secure
computability in these special cases.
Furthermore, for the cases with single-letter characterizations, the aforementioned
heuristic interpretation of $R^*$ is made precise (see the remark following Lemma \ref{l_11}
below).

The algorithms for exact function computation by multiple parties, without secrecy requirements, were first 
considered in \cite{Yao79}, and have since been studied extensively (cf. e.g., \cite{Gal88, GriKum05, KusNis97}). 
An information-theoretic version with asymptotically accurate (in observation length)
function computation was considered in \cite{OrlRoc01, MaIswGup09}.
The first instance of the exact function computation problem with secrecy appears in \cite{OrlEl84}. 
A basic version of the secure computation problem studied here was introduced in \cite{Tya10, Tya11}; 
\cite{Cha11} gives an alternative proof of the results in \cite{Tya10, Tya11}.

The problem of secure computing for multiple functions is
formulated in the next section, followed by our results in
section \ref{s_results}. The proofs are given in sections \ref{s_4} and \ref{s_5}. The 
final section discusses alternative forms of the necessary conditions.
 \vspace{0.2cm}

{\it Notation.} The set $\{1,..., m\}$ is denoted by $\cM$. For $i
< j$, denote by $[i,j]$ the set $\{i, ..., j\}$. Let $X_1,...,
X_m$, $m\geq 2$, be rvs taking values in finite sets $\cX_1, ... ,
\cX_m$, respectively, and with a known probability mass function.
Denote by $X_\cM$ the collection of rvs $\left(X_1, ...,
X_m\right)$, and by $X_\cM^n=\left(X_{\cM,1}, ..., X_{\cM,n}\right)$ the
$n$ independent and identically distributed (i.i.d). repetitions
of the rv $X_\cM$. For a subset $\cA$ of $\cM$, denote by $X_\cA$ the rvs
$\left(X_i, i \in \cA\right)$. Given $R_i \geq 0$, $1\leq i \leq m
$, let $R_\cA$ denote the sum $\sum_{i \in \cA}R_i$. Denote the cardinality of the range-space of an rv $U$
by $\|U\|$.

Finally, for $0 < \epsilon < 1$, an rv $U$ is
$\epsilon$-recoverable from an rv $V$ if there exists a function
$g$ of $V$ such that $\bPr{U = g(V)} \geq 1 - \epsilon$.

\section{Problem formulation}\label{s_formulation}

We consider a multiterminal source model for function computation
using public communication, with a confidentiality requirement. This basic
model was introduced in \cite{CsiNar04} in a separate context of SK
generation with public transaction. Terminals $1, \dots, m$
observe, respectively, the sequences $X_1^n, \ldots, X_m^n$ of
length $n$. For $0 \leq i \leq m$, let $g_i:\cX_\cM \rightarrow
\cY_i$ be given mappings, where the sets $\cY_i$ are finite.
Further, for  $0 \leq i \leq m$ and $n \geq 1$, the
(single-letter) mapping $g_i^n : \cX_\cM^n \rightarrow \cY_i^n$ is
defined by
\begin{align}\nonumber
g_i^n(x_\cM^n) &=(g_i(x_{11}, \ldots, x_{m1}), \ldots, g_i(x_{1n},
\ldots, x_{mn})), \\ \nonumber x_\cM^n &= (x_1^n, \ldots, x_m^n)
\in \cX_\cM^n.
\end{align}
For convenience, we shall denote the rv
$g_i^n\left(X_\cM^n\right)$ by $G_i^n, n \geq 1$, and, in
particular, $G_i^1 = g_i\left(X_\cM\right)$ simply by $G_i$.

Each terminal $i \in \cM$ wishes to compute the function $g_i^n(x^n_\cM)$,
without revealing $g_0^n(x^n_\cM)$, $x^n_\cM \in \cX^n_\cM$. To
this end, the terminals are allowed to communicate over a
noiseless public channel, possibly interactively in several
rounds.

\begin{definition}\label{d_PubComm}
An {\it $r$-rounds interactive communication protocol} consists of
mappings
$$f_{11}, ..., f_{1m}, ...., f_{r1}, ..., f_{rm},$$
where $f_{ij}$ denotes the communication sent by the $j$th node in
the $i$th round of the protocol; specifically, $f_{ij}$ is a
function of $X_j^n$ and the communication sent in the previous
rounds $\{f_{kl}: 1\leq k \leq i-1, l\in \cM\}$. Denote the rv
corresponding to the communication by
$$\mathbf{F} = F_{11}, ..., F_{1m},...., F_{r1},..., F_{rm},$$
noting that $\mathbf{F}= \mathbf{F}^{(n)}\left( X_\cM^n\right)$. The rate\footnote{All logarithms are with respect to the base $2$.} of $\mathbf{F}$ is $\frac{1}{n} \log \|\mathbf{F}\|$.
\end{definition}

\begin{definition}\label{d_SC1}
For $\ep_n > 0$, $n\geq 1$, we say that functions\footnote{ The
abuse of notation $g_\cM = \left(g_0, g_1, ..., g_m\right)$
simplifies our presentation.} $g_\cM = \left(g_0, g_1, ...,
g_m\right)$, with private function $g_0$,  are
$\ep_n$-\emph{securely computable} ($\ep_n$- SC) from observations
of length $n$, and public communication $\mathbf{F} =
\mathbf{F}^{(n)}$, if 
\begin{itemize}
\item[(i)] $G_i^n$ is $\ep_n$- recoverable from $( X_i^n,
\mathbf{F})$ for every $i \in \cM$,
and

\item[(ii)] $\mathbf{F}$ satisfies the secrecy condition
\end{itemize}
\begin{align}
\nonumber \frac{1}{n}I\left(G_0^n \wedge \mathbf{F}\right) \leq
\epsilon_n.
\end{align}
\end{definition}
\begin{remark*}
The definition of secrecy here corresponds to ``weak
secrecy" \cite{AhlCsi93}, \cite{Mau93}. When our results have a
single-letter form, our achievability schemes for secure computing attain
``strong secrecy" in the sense of \cite{Mau94}, \cite{Csi96},
\cite{CsiNar04}. In fact, when we have a single-letter form, our proof in section \ref{s_4} 
yields ``strong secrecy" upon minor modification. 
\end{remark*}

\begin{figure}
\begin{center}
\epsfig{file=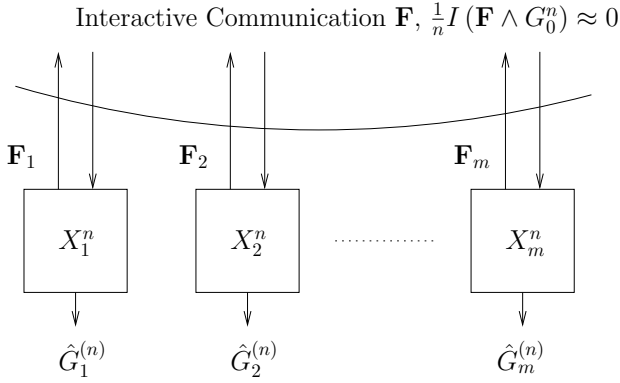, height=2in} \caption{Secure
computation of $g_1, ..., g_m$ with private function $g_0$}
\label{f_GSC}
\end{center}
\end{figure}

By definition, for $\epsilon_n$-SC functions $g_\cM$, the private
function $G_0$ is effectively concealed from an eavesdropper with
access to the public communication $\mathbf{F}$.
\begin{definition}\label{d_SC2}
For private function $g_0$, we say that functions $g_\cM$ are
\emph{securely computable} if $g_\cM$ are $\epsilon_n$- SC from
observations of length $n$
 and public communication
$\mathbf{F} = \mathbf{F}^{(n)}$, such that $\displaystyle \lim_n \ep_n = 0$.
Figure \ref{f_GSC} shows the setup for secure computing.
\end{definition}

In this paper, we give necessary and sufficient conditions for the
secure computability of certain classes of functions $g_\cM = \left(g_0, g_1,
..., g_m\right)$. 
The formulation in \cite{Tya11}, in which the terminals in a given subset 
$\cA$ of $\cM$ are required to compute (only) $g_0$ securely, is a special case
with
\begin{align}\label{e_c1}
g_i = \begin{cases}
g_0, &\quad i \in \cA,\\
\text{constant}, &\quad \text{otherwise.}
\end{cases}
\end{align}
It was shown in \cite{Tya11}
that (\ref{e_f1}) and (\ref{e_f2}) constitute, respectively,
necessary and sufficient conditions for the functions above to be
securely computable, with $R^*$ being the minimum rate of
interactive communication $\mathbf{F}$ that enables all the terminals in $\cM$ to
attain {\it omniscience} (see \cite{CsiNar04}), i.e., recover \emph{all} the data
$X_\cM^n$, using $\mathbf{F}$ and the {\it decoder side
information} $G_0^n$ given to the terminals in $\cM\setminus \cA$.
In fact, it was shown that when condition
(\ref{e_f2}) holds, it is possible to recover $X_\cM^n$ using
communication that is independent of $G_0^n$. 

%
%
 The guiding heuristic in this work is the following general principle,
 which is also consistent with the results 
 of \cite{Tya11}: 
 
 \emph{Conditions (\ref{e_f1}) and (\ref{e_f2}) constitute, respectively, the necessary 
 and sufficient conditions for functions $g_\cM = \left(g_0, g_1, ..., g_m\right)$ to be 
 securely computable, where $R^*$ is the infimum of the rates of interactive communication 
 $\mathbf{F}^\prime$ such that, for each $1\leq i \leq m$, the following hold simultaneously:}
\begin{itemize}
\item[(P1)] {\it $G_i^n$ is $\ep_n$-recoverable from $\left(X_i^n,
\mathbf{F}^\prime\right)$, and }

\item[(P2)] {\it $X_\cM^n$ is $\ep_n$-recoverable from $\left(X_i^n,
G_0^n, \mathbf{F}^\prime\right)$, i.e., terminals attain
omniscience, with $G_0^n$ as side information that is used only for decoding
(but is not used for the communication $\mathbf{F}^\prime$),}
\end{itemize}
{\it where $\ep_n \rightarrow 0$ as $n \rightarrow \infty$. }

Thus, (P1) and (P2) require any terminal computing $g_0$ to become omniscient, an
observation that was also made for the special case in \cite{Tya11}. The first condition 
(P1) above is straightforward 
and ensures the computability of the functions $g_1, ..., g_m$,
by the terminals $1, ..., m$, respectively. The omniscience 
condition (P2) facilitates the decomposition of total
entropy into mutually independent components that include 
the random values of the private function $G_0^n$ and the communication
$\mathbf{F}^\prime$. For the
specific case in (\ref{e_c1}), $R^*$ above has a single-letter
formula. In general, a single-letter expression for $R^*$ is not
known.

Our results, described in section \ref{s_results}, are obtained by
simple adaptations of this principle. Unlike \cite{Tya11}, our
conditions, in general, are not of a single-letter form.
Nevertheless, they provide a structural characterization of secure
computability. As an application, our results provide simple
conditions for secure computability in the following illustrative example.

\begin{example}\label{ex_1}
We consider the case of $m=2$ terminals that observe binary
symmetric sources (BSS) with underlying rvs $X_1,
X_2$ with joint pmf given by
\begin{align}\nonumber
\bPr{X_1 = 0, X_2 = 0} = \bPr{X_1 =1,  X_2 = 1} &=
\frac{1-\delta}{2},\\\nonumber \bPr{X_1 = 0, X_2 = 1} = \bPr{X_1 =
1, X_2 = 0} &= \frac{\delta}{2},
\end{align}
where $0< \delta < 1/2$. The results of this paper will allow us
to provide conditions for the secure computability of the four choices
of $g_0, g_1, g_2$ below; it will follow by Theorem \ref{t_1} that
functions $g_0, g_1, g_2$ are securely computable if
$$h(\delta) < \tau,$$
and conversely, if the functions above are securely computable,
then
$$h(\delta) \leq \tau,$$
where $h(\tau)= -\tau\log\tau -(1-\tau)\log(1-\tau)$, and the constant $\tau = \tau(\delta)$ depends on the choice of the function.
These characterizations are summarized in the next table. Denote the AND and the 
OR of two random bits $X_1$ and $X_2$ by
$X_1.X_2$ and $X_1 \oplus X_2$, respectively.
\vspace{0.3cm}

\begin{center}
\begin{tabular}{ | l | l | l | l |}\hline
$g_0$ &$ g_1$ & $g_2$ & $\tau$\\
 \hline
$X_1 \oplus X_2$ & $X_1 \oplus X_2$ & $X_1 \oplus X_2$ &$ 1/2
$\\\hline $X_1 \oplus X_2$ & $X_1 \oplus X_2$ & $\phi$ &
$1$\\\hline $X_1 \oplus X_2,\,\, X_1.X_2$ & $X_1 \oplus X_2,
\,\,X_1.X_2$ & $X_1.X_2$ & $ 2\delta/3 $\\\hline $X_1 \oplus X_2$ & $X_1
\oplus X_2$ &  $X_1.X_2$ & $2/3 $\\\hline
\end{tabular}

\end{center}
\vspace{0.3cm}

The results for the first two settings follow from \cite{Tya11}. The third and fourth results are new. In these
settings, terminal $1$ is required to recover the private
function; our results below show that the conditions for the
secure computability in these cases remain unchanged even if this
terminal is required to attain omniscience. Note that since $h(\delta)<1$
for all $0<\delta< 1/2$, there exists a communication protocol for securely computing the functions in the second setting.
By contrast, a secure computing protocol for the functions in the third setting does not exist for any $0<\delta< 1/2$, since 
$h(\delta) > 2\delta/3$. \qed
\end{example}

\section{Characterization of securely computable functions}\label{s_results}

In this section, we characterize securely computable functions for
three settings. Our necessary and sufficient conditions entail the
comparison of $H\left(X_\cM| G_0\right)$ with a rate $R^*$; the
specific choice of $R^*$ depends on the functions $g_\cM$.
Below we consider three different classes of functions
$g_\cM$. Although the first class is a special case of the second,
the two are handled separately as the more restrictive case
is amenable to simpler analysis. Furthermore, for $m=2$, the obtained
necessary and sufficient conditions for secure computability
take a single-letter form in the first case (see Corollary \ref{c_1}).

\vspace{0.3cm}

\noindent{\bf (1)} In the first class we consider, values of all the functions $g_1, ..., g_m$
must be kept secret. In addition, at least one of the terminals must 
compute all the functions $g_1, ..., g_m$. This case 
arises in distributed function computation over a network where all the
computed values are collated at a single sink node, and we are interested in
securing the collated function values. Alternatively,
denoting the function computed at the sink node by
the private function $g_0$, the computed functions 
$g_1, ..., g_m$ can be restricted to be functions of $g_0$.
Specifically, for $0 < m_0<m$, and for private function $g_0$,
let
\begin{align}\label{e_i01}
g_i = \begin{cases}
g_0,\quad &i \in \left[1,m_0\right],\\
g_i\left( g_0\right),\quad &i \in \left[m_0+1,m\right].
\end{cases}
\end{align}
\vspace{0.1cm}

\noindent{\bf (2)} The next case is a relaxation of the previous model
in that the restriction $g_i = g_i\left(g_0\right)$ for $i \in
\left[m_0+1, m\right]$ is dropped. 
For this general case, our analysis below implies roughly that 
requiring the terminals $\left[1, m_0\right]$ that compute the private function
$g_0$ to recover the entire data $X_\cM^n$ does not change the conditions
for secure computability, which is a key observation of this paper.
%
\vspace{0.2cm}

\noindent{\bf (3)} The last class of
problems we study is an instance of
\emph{secure multiterminal source coding}, which arises 
in the data download problems in sensor networks where
each node is interested in downloading the data observed 
by a subset of nodes. Specifically, we consider the
situation where each terminal wishes to recover some subset
$X_{\cM_i}^n$ of the sources where $\cM_i \subseteq\cM\setminus
\{i\}$, i.e., 
\begin{align}\label{e_i0}
g_i\left(X_\cM\right) = X_{\cM_i},\quad i \in \cM.
\end{align}
This last case appears to be disconnected from the previous 
two cases a priori. However, our characterizations of 
secure computability below have the same form for all 
cases above. Moreover, the same heuristic
principle, highlighted in (P1) and (P2), leads to 
a characterization of secure computability
in all three cases.

The necessary and sufficient conditions for 
secure computability are stated in terms of quantities
$R_i^*(g_\cM)$, $i = 1, 2, 3$, which are defined next. The subscript $i$
corresponds to case ($i$) above. In particular, the quantity $R^*$
corresponds to the minimum rate of communication needed for
an appropriate modification of the source-coding task in (P1), (P2).  
Below we give specific expressions for $R_i^*$, $i =1, 2, 3$, along with their
operational roles (for a complete description of this role see the 
sufficiency proof in Section \ref{s_4}). 

Denote by $\cR_1^*\left(g_\cM\right)$ 
the closure of the (nonempty) set of
pairs\footnote{The first term accounts for the rate of
the communication and the second term tracks the 
information about $G_0^n$ leaked by $\mathbf{F}$ (see (\ref{e_i04})) below}
$$\left(R_\mathbf{F}^{(1)}, \frac{1}{n}I\left(G_0^n\wedge \mathbf{F}\right)\right),$$
for all $n \geq 1$ and interactive communication $\mathbf{F}$,
where
\begin{align}\label{e_g1}
R_\mathbf{F}^{(1)} = \frac{1}{n}H(\mathbf{F}) + \frac{1}{n}\sum_{i
= m_0 +1}^m H\left(G_i^n| X_i^n, \mathbf{F}\right) +\inf R_\cM,
\end{align}
with the infimum taken over the rates $R_1, ..., R_m$ satisfying
the following constraints:
\begin{itemize}
\item[\bf (1a)] $\forall \cL \subsetneq \cM$, $\left[1,
m_0\right]\nsubseteq \cL$,
$$R_\cL \geq \frac{1}{n}H\left(X_\cL^n | X_{\cM\setminus \cL}^n, \mathbf{F}\right);$$
\item[\bf (1b)] $\forall \cL \subsetneq \cM$, $\left[1,m_0\right]
\subseteq \cL$,
$$R_\cL \geq \frac{1}{n}H\left(X_\cL^n | X_{\cM\setminus \cL}^n, G_0^n, \mathbf{F}\right).$$
\end{itemize}
The quantity $\inf_{n, \mathbf{F}}R^{(1)}_\mathbf{F}$ corresponds
to the solution of a multiterminal source coding problem. Specifically, it is
the infimum of the rates of interactive communication that
satisfy (P1) and (P2) above (see \cite[Theorem 13.15]{CsiKor11}, \cite{CsiNar04}).
\vspace{0.2cm}

Next, let $\cR_2^*\left(g_\cM\right)$ denote the closure of the set of pairs
$$\left(R_\mathbf{F}^{(2)}, \frac{1}{n}I\left(G_0^n\wedge \mathbf{F}\right)\right),$$
for all $n \geq 1$ and interactive communication $\mathbf{F}$,
where
\begin{align}\label{e_i02}
R_\mathbf{F}^{(2)} = \frac{1}{n}H(\mathbf{F}) + \inf
\left[R_{\left[m_0+1, m\right]}^\prime +R_\cM\right],
\end{align}
with the infimum taken over the rates $R_1, ..., R_m$ and
$R_{m_0+1}^\prime, ..., R_m^\prime$ satisfying the following
constraints:
\begin{itemize}
\item[\bf (2a)] $\forall \cL \subsetneq \cM$, $\left[1, m_0\right]
\nsubseteq \cL$,
$$R_\cL \geq \frac{1}{n}H\left(X_\cL^n | X_{\cM\setminus \cL}^n, \mathbf{F}\right);$$
\item[\bf (2b)] for $m_0< j \leq m$,
$$R_j^\prime \geq \frac{1}{n}H\left(G_j^n | X_j^n, \mathbf{F}\right);$$
\item[\bf (2c)] $\forall \cL \subseteq \cM, \left[1, m_0\right]\subseteq
\cL$, and $\cL^\prime \subseteq \left[m_0+1, m\right]$ with either $\cL
\neq \cM$ or $\cL^\prime \neq \left[m_0+1, m\right]$,
\begin{align}\nonumber
\hspace*{-0.2in} R_{\cL^\prime}^\prime + R_\cL \geq
\frac{1}{n}H\left(G_{\cL^\prime}^n, X_\cL^n| G^n_{\left[m_0+1,
m\right]\setminus \cL^\prime}, X_{\cM\setminus \cL}^n, G_0^n,
\mathbf{F}\right).
\end{align}
\end{itemize}
The quantity $\inf_{n, \mathbf{F}}R^{(2)}_\mathbf{F}$
corresponds to the solution of a multiterminal source coding problem, and is the infimum of the rates of interactive
communication $\mathbf{F}^\prime$ that satisfy (P1) and (P2)
above, and additionally satisfies: \vspace{0.1cm}

\begin{itemize}
\item[(P3)] {\it $X_\cM^n$ is $\ep_n$-recoverable from $\left(G_j^n,
G_0^n, \mathbf{F}^\prime\right)$, $m_0 < j\leq m$.}
\end{itemize}
\vspace{0.1cm}

\noindent This modification corresponds to the introduction of $m-
m_0$ dummy terminals, with the $j$th dummy terminal observing
$G_j^n$, $m_0< j \leq m$ (see section \ref{s_Disc}); the dummy terminals can be
realized by a terminal $i$ in $\left[1, ..., m_0\right]$ that recovers
$X_\cM^n$ from $\left(X_i^n, \mathbf{F}\right)$. The conditions (P2) and (P3) above correspond to the
omniscience at the terminals in the extended model, with $G_0^n$
provided as side information only for decoding.

Finally, denote by
$\cR_3^*\left(g_\cM\right)$ the closure of the set of pairs
$$\left(R_\mathbf{F}^{(3)}, \frac{1}{n}I\left(G_0^n\wedge \mathbf{F}\right)\right),$$
for all interactive communication $\mathbf{F}$, where
\begin{align}\label{e_i03}
R_\mathbf{F}^{(3)} = \frac{1}{n}H(\mathbf{F}) +\inf R_\cM,
\end{align}
with rates $R_1, ..., R_m$ satisfying the following constraints:
\begin{itemize}
\item[\bf (3a)] For $1\leq i \leq m$,   $\forall \, \cL\subseteq \cM_i \subseteq \cM\setminus \{i\}$,
 $$R_\cL \geq \frac{1}{n}H\left(X_\cL^n| X_{\cM_i\setminus \cL}^n, X_i^n, \mathbf{F}\right);$$
\item[\bf (3b)] $\forall \cL \subsetneq \cM$,
$$R_\cL \geq \frac{1}{n}H\left(X_\cL^n| X_{\cM\setminus \cL}^n, G_0^n, \mathbf{F}\right).$$
\end{itemize}
As before, the quantity $\inf_{n, \mathbf{F}}R^{(3)}_\mathbf{F}$ corresponds
to the infimum of the rates of interactive communication that
satisfy (P1) and (P2) above.

Our main result below characterizes securely computable functions
for the three settings above.
\begin{theorem}\label{t_1}
For $i = 1,2,3$, with functions $g_0, g_1, ..., g_m$ as in the
case ($i$) above, the functions $g_\cM$ are securely computable if
the following condition holds:
\begin{align}\label{e_g2}
H\left(X_\cM | G_0\right) > R_i^*\left(g_\cM\right).
\end{align}
Conversely, if the functions above are securely computable, then
\begin{align}\label{e_i00}
H\left(X_\cM | G_0\right) \geq R_i^*\left(g_\cM\right),
\end{align}
where
\begin{align}\label{e_i04}
R^*_i\left(g_\cM\right) = \inf_{(x, 0) \in \cR_i^*\left(g_\cM\right)} x, \quad i =1,2,3.
\end{align} 
\end{theorem}
\begin{remark*}
Although the first setting above is a special case of the second, it is unclear if for $g_\cM$
in (\ref{e_i01}) the quantities $R_1^*(g_\cM)$ and $R_2^*(g_\cM)$ are identical (also, see Section \ref{s_Disc}). In general, 
the multi-letter characterizations of secure computability of $g_\cM$ above can have 
different forms. 
For case (1) with $m=2$, Corollary $\ref{c_1}$ below provides a single-letter formula 
for $R_1^*(g_\cM)$. However, a similar single-letter formula for $R_2^*(g_\cM)$ is
not known.
\end{remark*}
Theorem \ref{t_1} affords the following heuristic interpretation. The quantity $H\left(X_\cM | G_0\right)$ represents 
the maximum rate of randomness in $X_\cM^n$ that is (nearly) independent of $G_0^n$. On the other hand, $R_i^*\left(g_\cM\right)$ is an appropriate rate of communication for the computation of $g_\cM$; we show that latter being less than $H\left(X_\cM | G_0\right)$ guarantees the secure computability of $g_\cM$.

Although the characterization in Theorem \ref{t_1} is not 
of a single-letter form, the following result provides a sufficient
condition for obtaining 
such forms.
Denote
by $R_{\text{constant}}^{(i)}$, $i= 1, 2, 3$, the quantity
$R_{\mathbf{F}}^{(i)}$ for $\mathbf{F} = \text{constant}$.

\begin{lemma}\label{l_11}
For case $(i)$, $i = 1,2,3$, if for all $n \geq 1$ and interactive
communication $\mathbf{F}$
\begin{align}\label{e_b1}
R_{\mathbf{F}}^{(i)} \geq
R_{\text{constant}}^{(i)},
\end{align}
then $R^*_i\left(g_\cM\right) = R_{\text{constant}}^{(i)} =\inf_{n,
\mathbf{F}}R_\mathbf{F}^{(i)}$.
\end{lemma}
The proof is a simple consequence of the definition of $R^*_i\left(g_\cM\right)$ in (\ref{e_i04}).
Note that $R_{\text{constant}}^{(i)}$ has a single-letter form. 
\begin{remark*}
As mentioned before, the quantity $\inf_{n,
\mathbf{F}}R_\mathbf{F}^{(i)}$ is the infimum of the rates of interactive communication that satisfies (P1), (P2) for $i = 1, 3$, and satisfies (P1)-(P3) for $i = 2$. Thus, when the conditions of Lemma \ref{l_11} hold, we have from Theorem \ref{t_1} that $g_\cM$ are securely computable if $$H\left(X_\cM | G_0\right) > R_{\text{constant}}^{(i)},$$ and if $g_\cM$ are securely computable then 
$$H\left(X_\cM | G_0\right)\geq R_{\text{constant}}^{(i)},$$
where $R_{\text{constant}}^{(i)}$ is the minimum rate of communication that satisfies (P1), (P2) for $i = 1, 3$, and satisfies (P1)-(P3) for $i = 2$. 
\end{remark*}

As a consequence of Lemma \ref{l_11}, we obtain below a single-letter
characterization of securely computable functions, with $m=2$, in a special
case; the following lemma, which is a special case of \cite[Lemma
B.1]{CsiNar08} (see also \cite[Theorem 1]{MadTet10}), is instrumental to our proof.
\begin{lemma}\label{l_1}
Let $m =2$. For an interactive communication $\mathbf{F}$, we have
$$H(\mathbf{F}) \geq H\left( \mathbf{F}| X_1^n\right) + H\left( \mathbf{F}| X_2^n\right).$$
\end{lemma}

We next consider case ($1$) for two terminals.
\begin{corollary}\label{c_1}
For $m=2$, for functions $g_0, g_1, g_2$ with $g_1 = g_0$  and
$g_2 = g_2\left( g_0\right)$, we have
\begin{align}\label{e_a2}
R_1^*\left(g_\cM\right) =  H\left(X_2| X_1\right) + H\left(G_2| X_2\right) +
H\left(X_1| X_2, G_0\right).
\end{align}
\end{corollary}
{\it Proof:} The constraints (1a) and (1b) satisfied by rates
$R_1, R_2$ in the definition of $R_{\mathbf{F}}^{(1)}$ are
\begin{align}\nonumber
R_2&\geq \frac{1}{n}H\left(X_2^n | X_1^n,
\mathbf{F}\right),\\\nonumber R_1&\geq \frac{1}{n}H\left(X_1^n |
X_2^n, G_0^n,\mathbf{F}\right),
\end{align}
which further yields
\begin{align}
\nonumber
R_{\mathbf{F}}^{(1)} &=  \frac{1}{n}\left[H\left(\mathbf{F}\right)
+ H\left(G_2^n | X_2^n,
\mathbf{F}\right)
\right.\\\label{e_e1}&\qquad \left.
+ H\left(X_2^n
| X_1^n, \mathbf{F}\right) + H\left(X_1^n | X_2^n, G_0^n,
\mathbf{F}\right)\right].
\end{align}
Thus, $R_{\text{constant}}^{(1)}$ equals the term on the right
side of (\ref{e_a2}). Upon manipulating the expression for $R_{\mathbf{F}}^{(1)}$ above, we get
\begin{align}
\nonumber
R_\mathbf{F}^{(1)} 
&= \frac{1}{n}\left[ H(\mathbf{F}) -  H\left(\mathbf{F}| X_1^n\right) -  H\left(\mathbf{F}| X_2^n, G_0^n\right)
\right.
\\ \label{e_a3} &\qquad \left.
- I\left(G_2^n \wedge \mathbf{F}| X_2^n\right)\right] + R_{constant}^{(1)}.
\end{align}
Further, since $H\left(G_2| G_0\right) = 0$, it holds that
$$I\left(G_2^n \wedge \mathbf{F}| X_2^n\right) \leq I\left(G_0^n \wedge \mathbf{F}| X_2^n\right),$$
which along with (\ref{e_a3}) yields
\begin{align}\nonumber
R_\mathbf{F}^{(1)}&\geq \frac{1}{n}\bigg[ H(\mathbf{F}) -
H\left(\mathbf{F}| X_1^n\right) - H\left(\mathbf{F}|
X_2^n\right)\bigg] + R_{constant}^{(1)} \\\nonumber &\geq
R_{constant}^{(1)},
\end{align}
where the last inequality follows from Lemma \ref{l_1}. The result
then follows from Lemma \ref{l_11}. \qed \vspace{0.2cm}

We next derive simple conditions for secure computability for the
BSS in Example \ref{ex_1}
\begin{example}\label{ex_2}
Consider the setup of Example \ref{ex_1}, with $g_0 = g_1 =
X_1\oplus X_2, X_1.X_2$ and $g_2 = X_1.X_2$. By Corollary
\ref{c_1} and the observation $H\left(G_2|X_2\right) =
h(\delta)/2$, we get $R_1^*\left(g_\cM\right) = 3h(\delta)/2$.
Since $H\left(X_1, X_2\mid G_0\right) = H\left(X_1,X_2\mid X_1\oplus X_2\right) - H\left(X_1.X_2\mid X_1\oplus X_2\right) = \delta$, the characterization of
secure computability claimed in Example \ref{ex_1} follows from
Theorem \ref{t_1}. \qed
\end{example}

\begin{example}\label{ex_3}
In the setup of Example \ref{ex_1}, consider $g_0 = g_1 =
X_1\oplus X_2$ and $g_2 = X_1.X_2$. This choice of $g_0, g_1, g_2$
is an instance of case ($2$) above. For an
interactive communication $\mathbf{F}$, 
the constraints (2a), (2b), (2c) in the definition of $R_\mathbf{F}^{(2)}$, upon simplification, reduce to
\begin{align}\nonumber
R_1&\geq \frac{1}{n}H\left(X_1^n | X_2^n, G_0^n, G_2^n, \mathbf{F}\right),\\\nonumber
R_2&\geq \frac{1}{n}H\left(X_2^n | X_1^n, \mathbf{F}\right),\\\nonumber
R_1+R_2 &\geq  \frac{1}{n}H\left(X_1^n, X_2^n| G_0^n, G_2^n, \mathbf{F}\right),\\\nonumber
R_2^\prime &\geq \frac{1}{n}H\left(G_2^n | X_2^n, \mathbf{F}\right).
\end{align}
Therefore, $\inf \left[R_1 + R_2 + R_2^\prime\right]$ with $R_1, R_2, R_2^\prime$ satisfying (2a), (2b), (2c), is given by
\begin{align}
\nonumber
&\frac{1}{n}\bigg[H\left(X_1^n| X_2^n, G_0^n, G_2^n, \mathbf{F}\right)
\\\nonumber 
&\quad +\max\left\{H\left(X_2^n| G_0^n, G_2^n, \mathbf{F}\right), H\left(X_2^n | X_1^n, \mathbf{F}\right)\right\}
\\\nonumber 
&\quad +  H\left(G_2^n | X_2^n,\mathbf{F}\right)
\bigg],
\end{align}
which further gives
\begin{align}
\nonumber
R_\mathbf{F}^{(2)} &=\frac{1}{n}\bigg[H(\mathbf{F}) + H\left(X_1^n| X_2^n, G_0^n, G_2^n, \mathbf{F}\right) 
\\\nonumber &  \quad + \max\left\{H\left(X_2^n| G_0^n, G_2^n, \mathbf{F}\right), H\left(X_2^n | X_1^n, \mathbf{F}\right)\right\}  
\\\label{e_a6} & \quad + H\left(G_2^n | X_2^n,\mathbf{F}\right)\bigg].
\end{align}
It follows from $H\left(X_1^n | X_2^n, G_0^n, G_2^n,
\mathbf{F}\right)=0$ that
\begin{align}\nonumber
R_{constant}^{(2)} &= H\left(G_2| X_2\right) 
\\\nonumber &\qquad +
\max\left\{H\left(X_2| G_0,G_2\right),H\left(X_2|
X_1\right)\right\}\\\label{e_c2} &= \frac{h(\delta)}{2} +
\max\left\{\delta, h(\delta)\right\} = \frac{3}{2}h(\delta),
\end{align}
as $h(\delta)> \delta$ for $0< \delta <1/2$.

Next, note from (\ref{e_a6}) that for any interactive
communication $\mathbf{F}$
\begin{align}\nonumber
R_\mathbf{F}^{(2)} &\geq
\frac{1}{n}\left[H(\mathbf{F}) + H\left(X_2^n| X_1^n, \mathbf{F}\right)+ H\left(G_2^n| X_2^n, \mathbf{F}\right)
\right] \\\nonumber
&=
\frac{1}{n}\left[H(\mathbf{F}) + H\left(X_2^n| X_1^n\right) \right.
\\\nonumber 
&\qquad\left. - H\left(\mathbf{F}| X_1^n\right) 
+ H\left(G_2^n, \mathbf{F}| X_2^n\right) - H\left(\mathbf{F}| X_2^n\right)\right]
\\\nonumber
&\geq \frac{1}{n}\left[H(\mathbf{F}) - H\left(\mathbf{F}| X_1^n\right) -
H\left(\mathbf{F}| X_2^n\right)\right]
\\\nonumber &\qquad + H\left(G_2| X_2\right) + H\left(X_2| X_1\right)
\\\label{e_c3}
&\geq H\left(G_2| X_2\right) + H\left(X_2| X_1\right) = \frac{3}{2}h(\delta),
\end{align}
where the last inequality above follows from Lemma \ref{l_1}. The
characterization in Example \ref{ex_1} follows from (\ref{e_c2}),
(\ref{e_c3}), and $H\left(X_1, X_2| G_0\right) = 1$, using Lemma
\ref{l_11} and Theorem \ref{t_1}. \qed
\end{example}
\section{Proof of sufficiency in Theorem \ref{t_1}}\label{s_4}

{\it Sufficiency of (\ref{e_g2}) for $i=1$:} We propose a two step protocol for 
securely computing $g_0, g_1, ..., g_m$. In the first step, for sufficient large $N$, the terminals 
$\left[1, m_0\right]$ ($g_0$-seeking terminals) attain omniscience, using an interactive communication 
$\mathbf{F}^{\prime \prime} = \mathbf{F}^{\prime \prime} \left(X_\cM^N\right)$ that satisfies 
\begin{align}\label{e_g3}
\frac{1}{N}I\left(G_0^N\wedge \mathbf{F}^{\prime \prime} \right)\leq \epsilon,
\end{align}
where $\ep > 0$ is sufficiently small. Next, upon attaining omniscience, one of the terminals in 
$\left[1, m_0\right]$ computes the following for $m_0 < j \leq m$:
\begin{itemize}
\item[(i)] Slepian-Wolf codewords $\hat{F}_j = \hat{F}_j\left(G_j^N\right)$ of appropriate rates $R_j^\prime$ for a recovery of $G_j^N$
 by a decoder with the knowledge of $X_j^N$ and previous communication $\mathbf{F}^{\prime\prime}$, and
\item[(ii)] the rvs $K_j = K_j\left(X_j^N\right)$ of rates $R_j^\prime$ that satisfy:
\begin{align}\label{e_g4}
\left|\frac{1}{N}H\left(K_j\right) - R_j^\prime\right| &\leq \epsilon, \\\label{e_g5}
\frac{1}{N}I\left(K_j \wedge G_0^N, \mathbf{F}^{\prime \prime}, \left\{K_l \oplus \hat{F}_l\right\}_{m_0< l \leq j-1}\right) &\leq \epsilon.
\end{align}
\end{itemize}
Note that $K_j \oplus \hat{F}_j$ denotes the encrypted version of the Slepian-Wolf code $\hat{F}_j$, encrypted with a one-time pad using the secret key (SK) $K_j$. Thus, terminal $j$, with the knowledge of $K_j$, can recover $\hat{F}_j$ from $K_j \oplus \hat{F}_j$, and hence can recover $G_j^N$. The operation $K_j \oplus \hat{F}_j$ is valid since the SK $K_j$ has size greater than 
$\|\hat{F}_j\|$.
Furthermore, we have from (\ref{e_g3}) and (\ref{e_g5}) that 
\begin{align}\nonumber
&\frac{1}{N}I\left(G_0^N \wedge \mathbf{F}^{\prime \prime}, \left\{K_j \oplus \hat{F}_j\right\}_{m_0< j \leq m}\right)
\\\nonumber &\leq 
\frac{1}{N}I\left(G_0^N \wedge \left\{K_j \oplus \hat{F}_j\right\}_{m_0< j \leq m} \mid \mathbf{F}^{\prime \prime}\right) + \ep
\\\nonumber &\leq \sum_{j = m_0+1}^m \frac{1}{N}\left[\log\| K_j \oplus \hat{F}_j\| \right.
\\\nonumber &\quad - \left. H\left(K_j \oplus \hat{F}_j \mid \mathbf{F}^{\prime \prime}, \left\{K_i \oplus \hat{F}_i\right\}_{m_0 < i \leq j-1}, G_0^N\right)\right] + \epsilon
\\\nonumber &\leq \sum_{j = m_0+1}^m \frac{1}{N}\bigg[H\left( K_j \right)
%
\\\nonumber
&\quad -  H\left(K_j \oplus \hat{F}_j \mid \mathbf{F}^{\prime \prime}, \left\{K_i \oplus \hat{F}_i\right\}_{m_0 < i \leq j-1}, G_0^N\right)\bigg] + 2\epsilon
\\\nonumber
&= \sum_{j = m_0+1}^m \frac{1}{N}\bigg[H\left( K_j \right)
\\\label{e_h1ii}
&\quad - H\left(K_j \mid \mathbf{F}^{\prime \prime}, \left\{K_i \oplus \hat{F}_i\right\}_{m_0 < i \leq j-1}, G_0^N\right)\bigg] + 2\epsilon
\\\nonumber &\leq 3m\ep,
\end{align}
where the third inequality above uses (\ref{e_g4})
and the last inequality follows from (\ref{e_g5}). The equality 
in (\ref{e_h1ii}) follows from the fact that $\hat{F}_j = \hat{F}_j\left(G_j^N\right)$ is a function of $G_0^N$, since $G_j$ is a function of $G_0$. We note that this is the only place in the proof where the functional relation between $G_j$ and $G_0$ is used. 

Thus, the communication $\left(\mathbf{F}^{\prime\prime}, K_j \oplus \hat{F}_j, m_0 < j \leq m\right)$ constitutes the required secure computing protocol for $g_{\cM}$. It remains to show the existence of $\mathbf{F}^{\prime \prime}$ and $K_j$, $m_0< j \leq m$ that satisfy (\ref{e_g3})-(\ref{e_g5}).

Specifically, when (\ref{e_g2}) holds for $i =1$, we have from the definition of $R_1^*\left(g_\cM\right)$ in (\ref{e_i04}) that for all $0< \ep \leq \ep_0$ ($\ep_0$ to be specified later), there exists $n\geq 1$ and interactive communication $\mathbf{F} = \mathbf{F}\left(X_\cM^n\right)$ such that 
\begin{align}\label{e_h2}
\frac{1}{n}I\left(G_0^n\wedge \mathbf{F}\right) < \epsilon,
\end{align}
and 
$$R_\mathbf{F}^{(1)} \leq R_1^*\left(g_\cM\right) + \frac{\epsilon}{2},$$
where $R_\mathbf{F}^{(1)}$ is as in (\ref{e_g1}). This further implies that there exist $R_1, ..., R_m$ satisfying (1a) and (1b) (for $\mathbf{F}$) such that 
\begin{align}\label{e_h3}
\hspace*{-0.1cm} \frac{1}{n}H(\mathbf{F}) + \frac{1}{n}\sum_{i = m_0+1}^mH\left(G_j^n\mid X_j^n, \mathbf{F}\right) +R_\cM
\leq R_1^*\left(g_\cM\right) + \ep.
\end{align}
Choosing 
$$\ep_0 < H\left(X_\cM \mid G_0\right) - R_1^*\left(g_\cM\right) - \delta,$$
for some $\delta	< H\left(X_\cM \mid G_0\right) - R_1^*\left(g_\cM\right) $, we get from (\ref{e_h2}) and (\ref{e_h3}) upon simplification:
\begin{align}\label{e_h4}
\frac{1}{n}\sum_{i = m_0+1}^mH\left(G_j^n\mid X_j^n, \mathbf{F}\right) +R_\cM +\delta
< \frac{1}{n} H\left(X_\cM^n \mid G_0^n, \mathbf{F}\right).
\end{align}
Next, for $k \geq 1$, denote by $\mathbf{F}^k = \left(\mathbf{F}_1, ..., \mathbf{F}_k\right)$ the i.i.d. rvs $\mathbf{F}_i = \mathbf{F}\left(X_{\cM, n(i-1)+1}, ..., X_{\cM, ni}\right)$, $1 \leq i \leq k$. Further, let $N = nk$. In Appendix A, we follow the approach in the proof of \cite[Theorem 5]{Tya11} and use (\ref{e_h4}) to show that for sufficiently large $k$ there exists an interactive communication $\mathbf{F}^\prime = \mathbf{f}^\prime\left(X_\cM^{nk}\right)$ of overall rate $R_\cM + \delta/2$ that satisfies the following:
\begin{align}\nonumber
X_\cM^{nk} \text{ is $\ep$-recoverable from $\left(X_i^N, \mathbf{F}^k, \mathbf{F}^\prime\right)$ for $1\leq i\leq m_0$},
\\\label{e_h5} \text{and from $\left(X_i^N,  \mathbf{F}^k, G_0^N, \mathbf{F}^\prime\right)$ for $m_0< i \leq m$,} 
\end{align}
and further,
\begin{align}\label{e_h6} 
\frac{1}{N} I\left(G_0^N, \mathbf{F}^k \wedge \mathbf{F}^\prime\right) < \ep.
\end{align}
The proposed communication $\mathbf{F}^{\prime \prime}$ comprises $\mathbf{F}^\prime, \mathbf{F}^k$, and condition  (\ref{e_g3}) follows from (\ref{e_h2}) and (\ref{e_h6}). Finally, we show the existence of $\hat{F}_j$ and $K_j$, $m_0 < j \leq m$, as above. From the Slepian-Wolf theorem \cite{SleWol73}, there exist rvs $\hat{F}_j = \hat{F}_j\left(G_j^N\right)$ of rates 
\begin{align}\label{e_h4'}
R_j^\prime\leq \frac{1}{N}H\left(G_j^N \mid X_j^N, \mathbf{F}^k\right) + \frac{\delta}{2m},
\end{align}
such that $G_j^N$ is $\ep$-recoverable from $\left(X_j^N, \mathbf{F}^k, \hat{F}_j\right)$, $m_0< j \leq m$, for $k$ sufficiently large. Suppose the rvs $K_{m_0+1}, K_{m_0+2}, ..., K_{j}$ of rates $R_{m_0+1}^\prime,R_{m_0+2}^\prime, ..., R_j^\prime$, respectively, satisfy (\ref{e_g4}) and (\ref{e_g5}) for some $j\leq m-1$. Denote by $\mathbf{F}^{\prime}(j)$ the communication $\left(\mathbf{F}^{\prime}, K_i\oplus \hat{F}_i, m_0< i \leq j\right)$ of rate $R^{(j)}$ that satisfies 
\begin{align}\label{e_h4''}
R^{(j)} \leq R_\cM + \frac{1}{N}\sum_{i = m_0+1}^j H\left(G_i^N\mid X_i^N, \mathbf{F}^k\right) + 
\delta
\end{align}
We have from (\ref{e_h4})-(\ref{e_h4''}) that
\begin{align}\label{e_h7}
R_{j+1}^\prime< \frac{1}{N} H\left(X_\cM^N\mid G_0^N, \mathbf{F}^k\right) - R^{(j)}.
\end{align}
Heuristically, since $X_\cM^N$ is recoverable from $\left(X_{j+1}^N, \mathbf{F}^k, \mathbf{F}^{\prime}\right)$, (\ref{e_h7}) gives
\begin{align}\nonumber
&\frac{1}{N} H\left(X_{j+1}^N\mid G_0^N, \mathbf{F}^k, \mathbf{F}^{\prime}(j)\right) 
\\\nonumber 
&\approx  
\frac{1}{N} H\left(X_\cM^N\mid G_0^N, \mathbf{F}^k\right) - \frac{1}{N}H\left(\mathbf{F}^{\prime}(j)\mid G_0^N, \mathbf{F}^k\right) \\\nonumber
&\geq \frac{1}{N} H\left(X_\cM^N\mid G_0^N, \mathbf{F}^k\right) - R^{(j)} \\\nonumber &> R_{j+1}^\prime.
\end{align}
Thus, a randomly chosen mapping $K_{j+1} = K_{j+1}\left(X_{j+1}^N\right)$ of rate $R_{j+1}^\prime$ is almost jointly-independent of $G_0^N, \mathbf{F}^k, \mathbf{F}^{\prime}(j)$ (see \cite{Csi96}). This argument is made rigorous using a version of the ``balanced coloring lemma" (see \cite{AhlCsi98}, \cite{CsiNar04}) given in Appendix B. Specifically, in Lemma \ref{l_B}, set $U = X_\cM^N$, $U^\prime = X_{j+1}^N$, $V = G_0^N, \mathbf{F}^k$, $h = \mathbf{F}^{\prime}(j)$, and 
\begin{align}\nonumber
\cU_0 &= \bigg\{x_\cM^N \in \cX^N_\cM : 
\\\nonumber & \qquad \quad x^N_\cM =
\psi_{j+1}\left(x^N_{j+1}, f^\prime\left(x^N_{\cM}\right),\mathbf{F}^k,
g^n_0\left(x_\cM^N\right)\right)
\bigg\},
\end{align}
for some mapping $\psi_{j+1}$, where $f^\prime\left(X_\cM^N\right) = \mathbf{F}^\prime$ is as in (\ref{e_h5}).
By the definition of $\mathbf{F}^\prime$,
\begin{align}\nonumber
\bPr{U \in \cU_0} \geq 1 - \ep,
\end{align}
so that condition (\ref{e_bound0})(i) preceding Lemma \ref{l_B} is
met.  Condition (\ref{e_bound0})(ii), too, is met from the definition of $\cU_0,  h$ and $V$.

Upon choosing
\begin{align}\nonumber
d = \exp\left[k\left(H\left(X_\cM^n | G_0^n, \mathbf{F}\right) -
\frac{n\delta}{2m}\right)\right],
\end{align}
in (\ref{e_bound1}), the hypotheses of Lemma \ref{l_B} are
satisfied for appropriately chosen $\lambda$, and for sufficiently large $k$. Then, by Lemma \ref{l_B}, with
\begin{align}\nonumber
r = \left\lceil\exp\left( NR_{j+1}^\prime\right)\right\rceil,\quad r' =
\left\lceil\exp\left(NR^{(j)}\right)\right\rceil,
\end{align}
and with $K_{j+1}$ in the role of $\phi$, it follows from (\ref{e_sin}) that there exists rv $K_{j+1} = K_{j+1}\left(X_{j+1}^N\right)$ that satisfies (\ref{e_g4}) and (\ref{e_g5}), for $k$ sufficiently large. The proof is completed upon repeating this argument for $m_0< j < m$. \qed

{\it Sufficiency of (\ref{e_g2}) for $i=2$:} The secure computing protocol for this case also consists of two stages. In the first stage, as before, the terminals $\left[1, m_0\right]$ ($g_0$-seeking terminals) attain omniscience, using an interactive communication $\mathbf{F}^{\prime \prime} = \mathbf{F}^{\prime \prime}\left(X_\cM^N\right)$. The second stage, too, is similar to the previous case and involves one of the omniscience-attaining terminals in $\left[1, m_0\right]$ transmitting communication 
$\hat{F}_j = \hat{F}_j\left(G_j^N\right)$ to the terminals $j$, for $m_0 < j\leq m$. However, the encryption-based scheme of the previous case is not applicable here; in particular, (\ref{e_h1ii}) no longer holds. Instead, the communication $\hat{F}_j$ now consists of the Slepian-Wolf codewords for $G_j^N$ given $X_j^N$, and previous communication $\mathbf{F}^{\prime \prime}$. We show below that if (\ref{e_g2}) holds, then there exist communication $\mathbf{F}^{\prime \prime}$ and $\hat{F}_j$, $m_0< j\leq m$, of appropriate rate such that the following holds:
$$\frac{1}{N}I\left(G_0^N \wedge \mathbf{F}^{\prime \prime}, \hat{F}_{m_0+1}, ..., \hat{F}_{m}\right) < \ep,$$
for sufficiently large $N$.

Specifically, when (\ref{e_g2}) holds for $i=2$, using similar manipulations as in the previous case we get that for all $0< \ep< \ep_0$, there exist interactive communication $\mathbf{F} = \mathbf{F}\left(X_\cM^n\right)$, and rates $R_1, ..., R_m, R_{m_0+1}^\prime, ..., R_m^\prime$ satisfying (2a)-(2c) (for $\mathbf{F}$) such that 
$$\frac{1}{n}I\left(G_0^n \wedge \mathbf{F}\right) < \frac{\epsilon}{2},$$
and 
\begin{align}\label{e_i1}
R_\cM + R^\prime_{\left[m_0+ 1, m\right]} +\delta < \frac{1}{n}H\left(X_\cM^n\mid G_0^n, \mathbf{F}\right),
\end{align}
with $\delta< H\left(X_\cM \mid G_0\right) - R_2^*\left(g_\cM\right) - \ep_0$; (\ref{e_i1}) replaces (\ref{e_h4}) in the previous case.

Next, for $N = nk$ consider $2m-m_0$ correlated sources $X_j^N$, $1\leq j \leq m$, and $G_j^N$, $m_0 < j \leq m$. Since $R_1, ..., R_m, R_{m_0+1}^\prime, ..., R_m^\prime$ satisfy (2a)-(2c), random mappings $F_j^\prime = F_j^\prime\left(X_j^N\right)$ of rates $R_j$, $1 \leq j \leq m$, and 
$F_{j+m-m_0}^\prime = F_{j+m-m_0}^\prime\left(G_j^N\right)$ of rates $R_j^\prime$, $m_0< j\leq m$ satisfy the following with high probability, for $k$ sufficiently large (see \cite[Lemma 13.13 and Theorem 13.14]{CsiKor11}):
\vspace*{0.1cm}

\begin{enumerate}
\item[(i)] for $1\leq i \leq m$, $X_\cM^{nk}$ is $\ep$-recoverable from $\left(F_1^\prime, ..., F_m^\prime, \mathbf{F}^k, X_i^{nk}\right)$;
\vspace*{0.1cm}

\item[(ii)] for $m_0 < j \leq m$, $G_j^{nk}$ is $\ep$-recoverable from $\left(F_{j+m-m_0}^\prime, \mathbf{F}^k, X_j^{nk}\right)$;
\vspace*{0.1cm}

\item[(iii)] for $m_0 < j \leq m$, $X_\cM^{nk}$ is $\ep$-recoverable from $\left(\mathbf{F}^\prime, \mathbf{F}^k, X_j^{nk}, G_0^{nk}\right)$ and from $\left(\mathbf{F}^\prime, \mathbf{F}^k, G_j^{nk}, G_0^{nk}\right)$, 
\end{enumerate}
\vspace*{0.1cm}
where $\mathbf{F}^k = \left(\mathbf{F}_1, ..., \mathbf{F}_k\right)$ are i.i.d. rvs $\mathbf{F}_i = \mathbf{F}\left(X_{\cM, n(i-1)+1}, ..., X_{\cM, ni}\right)$, $1 \leq i \leq k$.
It follows from (\ref{e_i1}) in a manner similar to the proof in Appendix A that there exist communication $F_j^\prime$, $1\leq j \leq 2m-m_0$ as above such that 
$$\frac{1}{nk}I \left(G_0^{nk}\wedge \mathbf{F}^\prime, \mathbf{F}^k\right) < \ep,$$
for sufficiently large $k$.

The first stage of the protocol entails transmission of $\mathbf{F}^k$, followed by the transmission of $F_1^\prime, ..., F_m^\prime$, i.e., $\mathbf{F}^{\prime\prime} = \left(\mathbf{F}^k, F_1^\prime, ..., F_m^\prime\right)$. The second stage of communication $\hat{F}_j$ is given by $F^\prime_{j+m-m_0}$, for $m_0 < j \leq m$.\qed
\vspace*{0.3cm}

{\it Sufficiency of (\ref{e_g2}) for $i=3$:} Using the definition of $R_3^*\left(g_\cM\right)$ and the manipulations above, the sufficiency condition (\ref{e_g2}) implies that for all $0 < \ep< \ep_0$, there exist interactive communication $\mathbf{F} = \mathbf{F}\left(X_\cM^n\right)$, and rates $R_1, ..., R_m$ satisfying (3a), (3b) (for $\mathbf{F}$) such that 
$$\frac{1}{n}I \left(G_0^n \wedge \mathbf{F}\right) < \frac{\epsilon}{2},$$
and 
\begin{align}\label{e_i2}
R_\cM +\delta < \frac{1}{n}H\left(X_\cM^n \mid G_0^n, \mathbf{F}\right),
\end{align}
for $\delta< H\left(X_\cM \mid G_0\right) - R_3^*\left(g_\cM\right) -\ep_0$. Denoting by $\mathbf{F}^k = \left(\mathbf{F}_1, ..., \mathbf{F}_k\right)$ the i.i.d. rvs $\mathbf{F}_i = \mathbf{F}\left(X_{n(i-1)+1}^{ni}\right)$, $1 \leq i \leq k$, it follows from (3a) and (3b) that for $N=nk$ the random mappings $F_i^\prime = F_i^\prime\left(X_i^{nk}\right)$ of rates $R_i$, $1\leq i \leq m$, satisfy the following with high probability, for $k$ sufficiently large (see \cite[Lemma 13.13 and Theorem 13.14]{CsiKor11}):
\begin{enumerate}
\item[(i)] for $i \in \cM$, $X_{\cM_i}^{nk}$ is $\ep$-recoverable from $\left(\mathbf{F}^\prime, \mathbf{F}^k, X_i^{nk}\right)$;

\item[(ii)] for $i \in \cM$, $X_\cM^{nk}$ is $\ep$-recoverable from $\left(\mathbf{F}^\prime, \mathbf{F}^k, X_i^{nk}, G_0^{nk}\right)$.
\end{enumerate}
From (\ref{e_i2}), the approach of Appendix A implies that there exist $F_i^\prime$, $i \in \cM$, as above such that 
$$\frac{1}{nk}I\left(G_0^{nk}\wedge \mathbf{F}^\prime, \mathbf{F}^k\right)< \ep,$$
for sufficiently large $k$. The interactive communication $\left(\mathbf{F}^\prime, \mathbf{F}^k\right)$ constitutes the protocol for securely computing $g_\cM$, where $g_i\left(X_\cM\right) = X_{\cM_i}, i \in\cM$.\qed

\section{Proof of Necessity in Theorem \ref{t_1}}\label{s_5}
{\it Necessity of (\ref{e_i00}) for $i=1$:} If functions $g_\cM$ are securely computable then there exists an interactive communication $\mathbf{F}$ such that $G_i^n$ is $\ep_n$-recoverable from $\left( X_i^n, \mathbf{F}\right)$, $i \in \cM$, and 
\begin{align}\label{e_j1}
\frac{1}{n}I\left(G_0^n \wedge \mathbf{F}\right) < \ep_n,
\end{align}
where $\ep_n \rightarrow 0$ as $n \rightarrow \infty$. 
It follows from the Fano's inequality that\footnote{The constants
$c_1, c_2, c_3, c_4$ depend only on $\log\|\cX_\cM\|$, $m$, $m_0$ (and not on $n$).}
\begin{align}\label{e_j2}
\frac{1}{n}H\left(G_i^n \mid X_i^n, \mathbf{F}\right) < c_1\ep_n,\qquad i \in \cM.
\end{align}
Using an approach similar to that in \cite{CsiNar04}, we have from (\ref{e_j1}):
\begin{align}\nonumber
&\frac{1}{n}H\left(X_\cM^n\right)
\\\nonumber &= \frac{1}{n}H\left(G_0^n, \mathbf{F}\right) + \frac{1}{n}H\left(X_\cM^n \mid G_0^n, \mathbf{F}\right)\\\label{e_j3'} &\geq
\frac{1}{n}H\left(G_0^n\right)  + \frac{1}{n}H\left(\mathbf{F}\right)   + \frac{1}{n}H\left(X_\cM^n \mid G_0^n, \mathbf{F}\right) -\ep_n,
\\\nonumber &=
\frac{1}{n}H\left(G_0^n\right)  + \frac{1}{n}H\left(\mathbf{F}\right)   + \frac{1}{n}\sum_{i=1}^mH\left(X_i^n \mid X^n_{[1, i-1]}, G_0^n, \mathbf{F}\right) \\\label{e_j3} &\hspace*{7cm}-\ep_n.
\end{align}
 Next, for $\cL \subsetneq \cM$, with $\left[1, m_0\right] \nsubseteq \cL$, we have 
 \begin{align}\nonumber
& \frac{1}{n	}H\left(X_\cL^n \mid X_{\cM\setminus \cL}^n, \mathbf{F} \right) 
 \\\nonumber &=  \frac{1}{n	}H\left(X_\cL^n \mid X_{\cM\setminus \cL}^n, G_0^n, \mathbf{F} \right) +  \frac{1}{n}H\left(G_0^n \mid X_{\cM\setminus \cL}^n, \mathbf{F} \right)\\\nonumber
 &\leq \frac{1}{n	}H\left(X_\cL^n \mid X_{\cM\setminus \cL}^n, G_0^n, \mathbf{F} \right) + c_1\ep_n,
 \end{align}
 where the last step follows from (\ref{e_j2}) and the assumption that $g_i = g_0$ for $i \in \left[1,m_0\right]$. Continuing with the inequality above, we get
 \begin{align}\nonumber
 &\frac{1}{n	}H\left(X_\cL^n \mid X_{\cM\setminus \cL}^n, \mathbf{F} \right) 
\\\label{e_j5} &\leq \frac{1}{n}\sum_{i\in \cL}\left[H\left(X_i^n \mid X_{[1, i-1]}^n, G_0^n, \mathbf{F} \right) + c_1\ep_n\right],
 \end{align}
Letting 
$$R_i = \frac{1}{n}H\left(X_i^n \mid X_{[1, i-1]}^n, G_0^n, \mathbf{F} \right) + c_1\ep_n, \quad i \in \cM,$$
by (\ref{e_j5}) $R_1, ..., R_m$ satisfy (1a) and (1b) for $\mathbf{F}$, whereby it follows from (\ref{e_j2}) and (\ref{e_j3}) that
\begin{align}\nonumber
&H\left(X_\cM \mid G_0 \right) 
\\\nonumber
 &\geq \frac{1}{n}H(\mathbf{F}) + \frac{1}{n}\sum_{i = m_0+1}^mH\left(G_i^n \mid X_i^n, \mathbf{F}\right) + R_\cM  - c_2\ep_n
\\\nonumber
&\geq  R_\mathbf{F}^{(1)} - c_2\ep_n,
\end{align}
where $\mathbf{F}$ satisfies (\ref{e_j1}). Taking the limit $n \rightarrow \infty$, and using the definition of $R_1^*\left(	g_\cM\right)$ we get $H\left(X_\cM \mid G_0\right) \geq R_1^*\left(g_\cM\right).$\qed

{\it Necessity of (\ref{e_i00}) for $i=2$:} If $g_\cM$ are securely computable, the approach above implies that there exists an interactive communication $\mathbf{F}$ satisfying (\ref{e_j1}) and (\ref{e_j2}) such that, with
\begin{align}\nonumber
R_i &= \begin{cases}
\frac{1}{n}H\left(X_i^n \mid X^n_{[1, i-1]}, G_0^n, \mathbf{F}\right) + c_1\ep_n, \quad 1\leq i \leq m_0,
\\ 
\\
\frac{1}{n}H\left(X_i^n \mid X^n_{[1, i-1]}, G^n_{\left[m_0+1, i-1\right]}, G_0^n, \mathbf{F}\right) + c_1 \ep_n,
\\\hspace*{5cm} \quad m_0< i \leq m,
\end{cases}\\\nonumber
R_j^\prime &= c_1\ep_n, \quad m_0 < j \leq m,
\end{align}
we have by (\ref{e_j3'}),
\begin{align}\nonumber
&H\left(X_\cM \mid G_0 \right)
\\\nonumber &\geq \frac{1}{n}H(\mathbf{F}) + \frac{1}{n}H\left(X_\cM^n \mid G_0^n, \mathbf{F}\right) -\ep_n\\\nonumber
&\geq \frac{1}{n}H(\mathbf{F}) + \frac{1}{n}\sum_{i=1}^{m_0}H\left(X_i^n \mid X^n_{\left[1, i-1\right]}, G_0^n, \mathbf{F}\right) 
\\\nonumber & \,\,+ \frac{1}{n}\sum_{i=m_0+ 1}^{m}H\left(X_i^n \mid X^n_{\left[1, i-1\right]}, G^n_{\left[m_0+1, i-1\right]}, G_0^n, \mathbf{F}\right)
-\ep_n\\\label{e_j7} 
&\geq \frac{1}{n}H(\mathbf{F}) + R_\cM + R^\prime_{\left[m_0+1, m\right]}- c_3\ep_n.
\end{align}
Furthermore, (\ref{e_j2}) and the assumption $g_i=g_0$, $1\leq i \leq m_0$, yield for $\left[1, m_0\right] \nsubseteq \cL \subsetneq \cM$ that
\begin{align}\nonumber
& \frac{1}{n	}H\left(X_\cL^n \mid X_{\cM\setminus \cL}^n, \mathbf{F} \right) 
\\\nonumber 
&\leq   \frac{1}{n	}H\left(X_\cL^n \mid X_{\cM\setminus \cL}^n, G_0^n, \mathbf{F} \right) + c_1\ep_n
\\\nonumber
 &\leq \sum_{i\in \cL,\, i\leq m_0}\left[\frac{1}{n}H\left(X_i^n \mid X_{[1, i-1]}^n, G_0^n, \mathbf{F} \right) + c_1\ep_n\right] +
 \\\nonumber
&  \sum_{i\in \cL,\, i >m_0}\left[\frac{1}{n}H\left(X_i^n \mid X^n_{[1, i-1]}, G^n_{\left[m_0+1, i-1\right]}, G_0^n, \mathbf{F}\right)  + c_1\ep_n\right] 
 &\qquad 
 \\\label{e_j8}
  &= R_\cL, 
 \end{align}
 and similarly, for $\left[1, m_0\right] \subseteq \cL \subseteq \cM$, $\cL^\prime \subseteq \left[m_0+1, m\right]$, with either $\cL \neq \cM$ or $\cL^\prime \neq \left[m_0+1, m\right]$ that 
 \begin{align}\nonumber
& \frac{1}{n}H\left(G_{\cL^\prime}^n, X_\cL^n| G^n_{\left[m_0+1,
m\right]\setminus \cL^\prime}, X_{\cM\setminus \cL}^n, G_0^n,
\mathbf{F}\right)
\\\nonumber &= \frac{1}{n}H\left(X_\cL^n| G^n_{\left[m_0+1,
m\right]\setminus \cL^\prime}, X_{\cM\setminus \cL}^n, G_0^n,
\mathbf{F}\right)
\end{align}
\begin{align}
\nonumber
&\leq \frac{1}{n}H\left(X_\cL^n \mid X_{\cM\setminus \cL}^n, G_0^n, \mathbf{F}\right)
\\\label{e_j9}
&\leq R_\cL + R^\prime_{\cL^\prime},
 \end{align}
Therefore, (\ref{e_j8}), (\ref{e_j2}) and (\ref{e_j9}) imply that $R_1, ..., R_m$, $R^\prime_{m_0}, ..., R^\prime_{m}$ satisfy (2a)-(2c) for $\mathbf{F}$, which along with (\ref{e_j7}) yields 
\begin{align}\nonumber
H\left(X_\cM \mid G_0 \right) 
\geq  R_\mathbf{F}^{(2)} - c_3\ep_n,
\end{align}
where $R_\mathbf{F}^{(2)}$ is as in (\ref{e_i02}), and $\mathbf{F}$ satisfies (\ref{e_j1}), which completes the proof of necessity (\ref{e_i00}) for $i =2$ upon taking the limit $n \rightarrow \infty$.
\qed

{\it Necessity of (\ref{e_i00}) for $i=3$:}
If the functions $g_\cM$ in (\ref{e_i0}) are securely computable then, as above, there exists an interactive communication $\mathbf{F}$ that satisfies (\ref{e_j1}) and (\ref{e_j2}). Defining
$$R_i = \frac{1}{n}H\left(X_i^n \mid X^n_{[1, i-1]}, G_0^n, \mathbf{F}\right) + c_1\ep_n, \quad i \in \cM,$$
similar manipulations as above yield
\begin{align}\label{e_j11}
H\left(X_\cM \mid G_0 \right) &\geq \frac{1}{n}H(\mathbf{F}) + R_\cM - c_4\ep_n.
\end{align}
Further, from (\ref{e_j2}) we get that $R_1, ..., R_m$ satisfy (3a) and (3b) for $\mathbf{F}$. It follows from (\ref{e_j11}) that
\begin{align}\nonumber
H\left(X_\cM \mid G_0 \right) 
\geq  R_\mathbf{F}^{(3)} - c_4\ep_n,
\end{align}
where $R_\mathbf{F}^{(2)}$ is as in (\ref{e_i03}), and $\mathbf{F}$ satisfies (\ref{e_j1}), which completes the proof of necessity (\ref{e_i00}) for $i =3$ as above.
\qed

\section{Discussion: Alternative necessary conditions for secure computability}\label{s_Disc}

The necessary condition (\ref{e_i00}) for secure computing given in section \ref{s_results} is in terms of quantities $R_\mathbf{F}^{(i)}$, $i =1, 2, 3$, defined in (\ref{e_g1}), (\ref{e_i02}), (\ref{e_i03}), respectively. As remarked before, for $i = 1, 3$, the quantity $\inf_\mathbf{F} R_\mathbf{F}^{(i)}$ is the infimum over the rates of interactive communication that satisfy conditions (P1) and (P2). However, this is not true for $i=2$. Furthermore, although $i=1$ is special case of $i=2$, it is not clear if the necessary condition (\ref{e_i00}) for $i=2$ reduces to that for $i=1$ upon imposing the restriction in (\ref{e_i01}). In this section, we shed some light on this baffling observation.

First, consider the functions $g_\cM$ in (\ref{e_c1}). For this choice of functions, denoting by $R^*_0$ the minimum rate of interactive communication that satisfies (P1) and (P2), the results in \cite{Tya11} imply that (\ref{e_f1}) constitutes a necessary condition for secure computability, with $R^* = R^*_0$. 

Next, consider an augmented model obtained by introducing a new terminal $m+1$ that observes rv $X_{m+1} = \tilde{g}\left(X_\cM\right)$ and seeks to compute $g_{m+1} = \emptyset$. Further, the terminal does not communicate, i.e., observation $X_{m+1}^n$ is available only for decoding. Clearly, secure computability in the original model implies secure computability in the new model. It follows from the approach of \cite{Tya11} that for the new model also, (\ref{e_f1}) constitutes a necessary condition for secure computability, with $R^*$ now being the minimum rate of interactive communication that satisfies (P1) and (P2) when terminal $m+1$ does not communicate; this $R^*$ is given by $$\max\{H\left(X_\cM \mid \tilde{g}(X_\cM), G_0\right), R^*_0\}.$$
Note that the new necessary condition (\ref{e_f1}) is
$$H\left(X_\cM \mid G_0\right) \geq R^*_0 = \max\{H\left(X_\cM \mid \tilde{g}(X_\cM), G_0\right), R^*_0\},$$
which is, surprisingly, same as the original condition 
$$H\left(X_\cM \mid G_0\right) \geq R^*_0.$$ 

Our necessary condition (\ref{e_i00}) for $i=2$ is based on a similar augmentation that entails introduction of $m- m_0$ new terminals observing $g_{m_0+1}\left(X_\cM\right), ..., g_{m}\left(X_\cM\right)$ (to be used only for decoding). Now, however, this modification may result in a different necessary condition.
\section*{Appendix A}
\setcounter{equation}{0}
\renewcommand{\theequation}{A\arabic{equation}}
\setcounter{theorem}{0}
\renewcommand{\thetheorem}{A\arabic{theorem}}

From (\ref{e_h4}), we have 
$$nR_\cM +\frac{\delta}{2}
<  H\left(X_\cM^n \mid G_0^n, \mathbf{F}\right),
$$
where $R_1, ..., R_m$ satisfy conditions (1a) and (1b).
For each $i$ and $R_i \geq 0$, consider
a (map-valued) rv $J_i$ that is uniformly distributed on the
family $\cJ_i$ of all mappings $\cX^{nk}_i \rightarrow \{1, \ldots,
\lceil\exp(knR_i)\rceil\},$ $i \in \cM$. The rvs $J_1, ..., J_m,
X^{nk}_\cM$ are taken to be mutually independent. 

Fix $\ep, \ep'$, with $\ep' > m\ep$ and $\epsilon + \epsilon' <
1$. It follows from the proof of the general source network coding
theorem \cite[Lemma 13.13 and Theorem 13.14]{CsiKor11} that for
all sufficiently large $k$,
\begin{align}\nonumber
&\mathtt{Pr}\bigg(\bigg\{j_\cM \in \cJ_\cM: X^{nk}_\cM \text{ is }
\ep\text{-recoverable from}
\\&\qquad \quad \left(X^{nk}_i,j_{\cM\setminus \{i\}}\left(X^{nk}_{\cM\setminus \{i\}}\right),
Z_i^k\right), i \in \cM\bigg\} \bigg)\label{e_A1}
\geq 1 - \ep,
\end{align}
where, for $i \in \cM$,
$$Z_i^k= \begin{cases} \mathbf{F}^k,  \quad j\in \left[1, m_0\right],\\
\left(\mathbf{F}^k, G_0^{nk}\right), \quad m_0 < j\leq m.
\end{cases}
$$
%
Below we shall establish that
\begin{align}\label{e_A2}
\bPr{\left\{j_\cM \in \cJ_\cM: \frac{1}{nk}I\left(j_\cM(X^{nk}_\cM) \wedge
G_0^{nk}, \mathbf{F}^k\right) \geq \ep\right\}} \leq \ep',
\end{align}
for all $k$ sufficiently large, to which end it suffices to show
that
\begin{align}\nonumber &
\mathtt{Pr}\bigg(\bigg\{j_\cM \in \cJ_\cM: 
\\\nonumber &\,\qquad \frac{1}{nk}I\left(j_i(X^{nk}_i) \wedge G_0^{nk}, \mathbf{F}^k,
j_{\cM\setminus \{i\}}\left(X^{nk}_{\cM\setminus \{i\}}\right)\right)
\geq \frac{\ep}{m}\bigg\}\bigg) 
\\\label{e_A3} &\leq \frac{\ep'}{m}, \quad i \in
\cM,
\end{align}
since
\begin{align}\nonumber
&I\left(j_\cM\left(X^{nk}_\cM\right) \wedge G_0^{nk}, \mathbf{F}^k\right) 
\\\nonumber &= \sum_{i=1}^m
I\left(j_i\left(X^{nk}_i\right)\wedge G_0^{nk}, \mathbf{F}^k \mid j_1\left(X^{nk}_1\right),
\ldots, j_{i-1}\left(X^{nk}_{i-1}\right)\right)\\\nonumber &\leq
\sum_{i=1}^m I\left(j_i\left(X^{nk}_i\right)\wedge G_0^{nk}, \mathbf{F}^k,
j_{\cM\setminus \{i\}}\left(X^{nk}_{\cM\setminus \{i\}}\right)\right).
\end{align}
Then it would follow from (\ref{e_A1}), (\ref{e_A2}), and
definition of $Z_\cM$ that
\begin{align}\nonumber
&\mathtt{Pr}\bigg(\bigg\{ j_\cM \in \cJ_\cM: X_\cM^{nk} \text{ is }
\ep\text{-recoverable from
}
\\\nonumber &\hspace*{1.5cm}\left(X^{nk}_i, Z_i^k,
j_{\cM\setminus \{i\}}\left(X^{nk}_{\cM\setminus \{i\}}\right)\right),
i \in \cM, \text{ and } \\\nonumber &\hspace*{1.5cm}
\frac{1}{nk}I\left(j_\cM(X^{nk}_\cM) \wedge
G_0^{nk}, \mathbf{F}^k\right)  < \ep\bigg\}\bigg)\geq 1 - \ep - \ep'.
\end{align}
This shows the existence of a particular realization $\mathbf{F}^\prime$ of
$J_\cM$ that satisfies (\ref{e_h5}) and (\ref{e_h6}).

It now remains to prove (\ref{e_A3}). 
Defining
\begin{align}\nonumber
&\tilde{\cJ}_i= \bigg\{j_{\cM\setminus \{i\}} \in
\cJ_{\cM\setminus \{i\}}:\,  X_\cM^{nk}\text{ is
}\ep\text{-recoverable from
}
\\\nonumber &\hspace*{3cm}\left(X^{nk}_i,Z_i^{k}, j_{\cM\setminus \{i\}}\left(X^{nk}_{\cM\setminus \{i\}}\right),
\right)\bigg\},
\end{align}
we have by (\ref{e_A1}) that $\bPr{J_{\cM\setminus \{i\}} \in
\tilde{\cJ}_i} \geq 1- \ep$. It follows that
\begin{align}\nonumber
&\mathtt{Pr}\bigg(\bigg\{j_\cM \in \cJ_\cM: 
\\\nonumber &\qquad \frac{1}{nk}I\left(j_i(X^{nk}_i) \wedge G_0^{nk}, \mathbf{F}^k,
j_{\cM\setminus \{i\}}\left(X^{nk}_{\cM\setminus \{i\}}\right)\right)
\geq \frac{\ep}{m}\bigg\}\bigg)\\\nonumber &\leq \ep +
\sum_{j_{\cM\setminus \{i\}} \in
\tilde{\cJ}_i}\bPr{J_{\cM\setminus \{i\}} = j_{\cM\setminus \{i\}}}p\left( j_{\cM\setminus \{i\}}\right),
\end{align}
since $J_i$ is independent of $J_{\cM\setminus \{i\}}$, where
$p\left( j_{\cM\setminus \{i\}}\right)$ is defined as
\begin{align}\nonumber
&\mathtt{Pr}\bigg(\bigg\{j_i \in \cJ_i : \\\nonumber &\quad
\frac{1}{nk}I\left(j_i(X^{nk}_i) \wedge G_0^{nk}, \mathbf{F}^k,
j_{\cM\setminus \{i\}}\left(X^{nk}_{\cM\setminus \{i\}}\right)\right)
\geq
\frac{\ep}{m}\bigg\}\bigg).
\end{align}
Thus, (\ref{e_A3}) will follow upon
showing that 
\begin{align}\label{e_A5}
p\left( j_{\cM\setminus \{i\}}\right) \leq \frac{\ep'}{m} - \ep,\quad
j_{\cM\setminus \{i\}} \in \tilde{\cJ}_i,
\end{align}
for all $k$ sufficiently large. Fix $j_{\cM\setminus \{i\}} \in
\tilde{\cJ}_i$. We take recourse to Lemma \ref{l_B} in Appendix B,
and set $U = X^{nk}_\cM$, $U' = X^{nk}_i, V = \left(G_0^{nk}, \mathbf{F}^k\right), h =
j_{\cM\setminus \{i\}}$, and
\begin{align}\nonumber
&\cU_0 = \bigg\{x_\cM^{nk} \in \cX^{nk}_\cM : x^{nk}_\cM =
\psi_i\bigg(x^{nk}_i,
j_{\cM\setminus \{i\}}\left(x^{nk}_{\cM\setminus \{i\}}\right), 
\\\nonumber &\hspace*{2.5cm}\mathbf{F}^k\left(x_\cM^{nk}\right), g_0^n\left(x_\cM^n\right)\mathbf{1}\left(m_0< i \leq m\right)\bigg)
\bigg\}
\end{align}
for some mapping $\psi_i$. By the definition of $\tilde{\cJ}_i$,
\begin{align}\nonumber
\bPr{U \in \cU_0} \geq 1 - \epsilon,
\end{align}
so that condition (\ref{e_bound0})(i) preceding Lemma \ref{l_B} is
met.  Condition (\ref{e_bound0})(ii), too, is met from the definition of $\cU_0, h$ and $V$.

Upon choosing
\begin{align}\nonumber
d = \exp\left[k\left(H\left(X_\cM^n | G_0^n, \mathbf{F}\right) -
\frac{\delta}{2}\right)\right],
\end{align}
in (\ref{e_bound1}), the hypotheses of Lemma \ref{l_B} are
satisfied, for appropriately chosen $\lambda$, and for sufficiently large $k$. Then, by Lemma \ref{l_B}, with
\begin{align}\nonumber
r = \left\lceil\exp\left( knR_i\right)\right\rceil,\quad r' =
\left\lceil\exp\left(knR_{\cM\setminus i}\right)\right\rceil,
\end{align}
and with $J_i$ in the role of $\phi$, (\ref{e_A5}) follows from (\ref{e_bc}) and (\ref{e_sin}). \qed

\section*{Appendix B}
\setcounter{equation}{0}
\renewcommand{\theequation}{B\arabic{equation}}
\setcounter{theorem}{0}
\renewcommand{\thetheorem}{B\arabic{theorem}}

Our proof of sufficiency in Theorem \ref{t_1} requires random mappings to satisfy certain ``almost independence" and ``almost uniformity" properties. The following version of the ``balanced coloring lemma" given in \cite{Tya11} constitutes the key step in the derivation of these properties. 

Consider rvs $U, U', V$ with values in finite sets $\cU, \cU',
\cV$, respectively, where $U'$ is a function of $U$, and a mapping
$h : \cU \rightarrow \{1, \ldots, r'\}$. For $0<\lambda < 1$, let
$\cU_0$ be a subset of $\cU$
such that

\noindent (i) $\bPr{U \in \cU_0} > 1 - \lambda^2$;

\noindent (ii) given the event $\{U \in \cU_0, h(U) = j, U' = u', V = v\}$,
there exists $u = u(u') \in \cU_0$ satisfying
\begin{align}\nonumber
&\bPr{U' = u' \mid h(U)=j, V=v, U \in \cU_0}
\\&= \bPr{U = u \mid
h(U) = j, V= v, U \in \cU_0} ,\label{e_bound0}
\end{align}
for $1\leq j \leq r'$ and $v \in \cV.$
Then the following holds.
\begin{lemma}\label{l_B} Let the rvs $U, U', V$ and the set $\cU_0$ be as above. Further, assume that
\begin{align}\label{e_bound1}
\bP{UV}{\left\{ (u,v) : \bPr{U = u\mid V =v} > \frac{1}{d}
\right\}} \leq \lambda^2.
\end{align}
Then, a randomly selected mapping $\phi: \cU' \rightarrow \{1,
\ldots, r\}$ fails to satisfy
\begin{align}&\nonumber
\sum_{j=1}^{r'} \sum_{v \in {\cal V}} \bPr{h(U) = j, V
=v}\times
\\&\quad\sum_{i=1}^{r}  \left|\sum_{\substack{u' \in \cU':\\
\,\phi(u')= i}} \bPr{U' = u' \mid h(U) = j, V = v}- \frac{1}{r}
\right| < 14 \lambda, \label{e_bc}
\end{align}
with probability less than $2rr'|{\cal
V}|\exp\left(-\frac{c\la^3d}{rr'}\right)$ for a constant $c > 0 $.
\end{lemma}
\begin{remark*}

Denoting by $s_{var}$ the left side of (\ref{e_bc}), it follows from \cite[Lemma 1]{CsiNar04} that
\begin{align}\nonumber
\log r - H(\phi(U)) + I(\phi(U)\wedge h(U), V) \leq s_{var}\log \frac{r}{s_{var}}.
\end{align}
Since the function $f(x) = x\log (r/x)$ is increasing for $0 < x < re$, it follows from (\ref{e_bc}) that  
\begin{align}
\log r - H(\phi(U)) + I(\phi(U)\wedge h(U), V)
 \leq 14\lambda\log \frac{|\cU|}{14\lambda} .
 \label{e_sin}
\end{align}
\end{remark*}
\section*{Acknowledgements}
The author would like to thank Prof. Prakash Narayan for many helpful discussions
on this work. His detailed comments on an earlier draft helped improve
this manuscript. 
%


\begin{IEEEbiographynophoto}
{Himanshu Tyagi} received the Bachelor of Technology degree in
electrical engineering and the Master of Technology degree in
communication and information technology, both from the Indian
Institute of Technology, Delhi, India in 2007. He is currently a
Ph.D. candidate at the University of Maryland, College Park, USA.
\end{IEEEbiographynophoto}

\end{document}